# Best practices for a proper evaluation and conversion of physical property equations in superconductors: the examples of WHH formulation, Bean model and other cases of interest

Chiara Tarantini[1,*], Evgeny F. Talantsev[2,3], Ilaria Pallecchi[4], Jens Hänisch[5], Jeffery L. Tallon[6], David C. Larbalestier[1,7]

[1] National High Magnetic Field Laboratory, Florida State University, Tallahassee, FL 32310, USA
[2] M.N. Miheev Institute of Metal Physics, Ural Branch, Russian Academy of Sciences, Ekaterinburg, 620108, Russia
[3] NANOTECH Centre, Ural Federal University, Ekaterinburg, 620002, Russia
[4] CNR-SPIN, Physics Department, Genova, 16146, Italy
[5] Institute for Technical Physics, Karlsruhe Institute of Technology, Eggenstein-Leopoldshafen, 76344, Germany
[6] Robinson Research Institute, Victoria University of Wellington, Lower Hutt 5046, New Zealand
[7] Departments of Mechanical Engineering and Materials Science & Engineering, FAMU-FSU College of Engineering, Florida State University, Tallahassee, FL 32310, United States of America

**Abstract**

In recent years there have been growing concerns about the proper evaluation of physical properties of superconductors, in particular for quantities extracted from magnetic characterizations. Errors can and often do occur due to the following issues: i) several measurement instruments still use Gaussian & cgs-emu units instead of the preferable International System (SI) units; ii) there are decades of valuable publications where, however, equations were expressed in Gaussian & cgs-emu or other unit systems or where constants were normalized to unity, which requires proper understanding and unit conversion in order to correctly evaluate the measured physical quantities; iii) the conversion between unit systems sometimes appears challenging and may not be properly performed. In this paper we will describe how to properly convert physical quantities relevant for the evaluation of magnetic and other properties focusing on the still most used unit systems, SI and Gaussian & cgs-emu. We will provide examples of how to properly verify and understand the physical formulae. We will include examples for the correct method to determine the critical current density $J_c$ of a superconductor from the measurement of its magnetic hysteresis loop through the Bean model, and the correct conversion to SI of the equations for $H_{c2}(T)$ according to the Werthamer-Helfand-Hohenberg (WHH) formulation. The goals of this paper are to make the readers aware of the unit conversion issue, to provide useful hands-on tools for proper conversion and to strongly encourage future exclusive use of the SI units and formulae.

## 1. Introduction

In recent years, uncertainty has grown concerning the correct evaluation of magnetic and other properties in superconducting materials. Identification of errors caused by unit conversion has led to paper withdrawals or corrections. Major concerns are in particular related to the proper evaluation of $J_c$ from magnetic measurements, $J_{c,mag}$, for which two possible sources of errors were identified (besides the typical uncertainties due to geometrical dimensions, sample orientation, device sensitivities etc.): (i) the unit conversion itself, due to the fact that many instruments still provide the magnetic moment in emu, and (ii) the correct Bean formula [1], [2] to be used for the calculation of $J_{c,mag}$ from the measurement of the magnetization hysteresis amplitude ($\Delta M = M_{decr.H} - M_{inc.H}$). A major impediment is the different formulae used in literature for the same geometry. These formulae sometimes do not specify the unit system, which creates uncertainty, or they provide potentially misleading information or are simply incorrect.

---

[*] Corresponding author: tarantini@asc.magnet.fsu.edu

This paper is based on the fundamental physical principle of dimensional homogeneity, which should be always complied with. Dimensional homogeneity requires that in a valid physical equation the dimensions (units) of added or subtracted terms must be the same, arguments of exponential, trigonometric and other functions are dimensionless, and the two sides of an equation must have identical units: this means that only variables with the same physical nature can be equated. This is often not the case in common formulae used for conversion of magnetic moments to magnetization in practical units. This makes unit analysis essential as a tool to verify the measure system as well as the validity of a formula, within a dimensionless factor, and to understand whether all appropriate constants have been made explicit. Moreover, an effective way to assure correct computation when using an equation is to substitute the variables with their numerical values AND their units so that, after simplification, dimensional correctness can be verified. In general, it is important to clarify early on the unit system of a certain formula, looking for hints that might easily identify it (for example, $\mu_0$ is used only in SI). Hints for omitted fundamental constants are the apparent lack of dimensional homogeneity. Very common in the study of superconductors is, for instance, the expression of vortex activation energies in Kelvin, which implies the omission of the Boltzmann constant $k_B$ (more appropriate is to express the activation energy as $U_0/k_B$ in units of [K]). While this example is well understood, not using clearly stated practical units in other cases may lead to confusion. One example is the Ginzburg-Levanyuk number $Gi$ for the fluctuation range, considered below. Since $Gi$ depends on powers of critical lengths and may vary over orders of magnitude between materials, using inappropriate formulae may lead to substantial hidden errors.

Because of our own experience in the difficulties of changing units, and to avoid proliferation of incorrect formulae, here we will first discuss common causes of errors in the unit conversion providing a table and describing the best practice to avoid such mistakes. Then, we provide a few examples of formulae important in superconductivity that have been, on occasions, reported incorrectly or not properly defined: we will consider the McMillan formula [3], the Ginzburg-Levanyuk number [4], [5] and the Werthamer-Helfand-Hohenberg (WHH) equations [6]. Finally, we will calculate $J_{c,mag}$ in SI units for a sample of rectangular cross-section in perpendicular field using the simplest Bean model scenario [1], [2], being explicit about how to properly convert the quantities into SI units. Since the cgs-emu units are still used widely because most popular magnetometers present their output in this way, we then show how to write a converted formula which allows direct use of cgs-emu quantities to evaluate $J_{c,mag}$.

Considering how easy it is to make mistakes in unit or formula conversion, it is thus absolutely pertinent to cite the final paragraph of Goldfarb's paper [7], which earlier also explored this territory. He cites and comments on an over-100-year-old paper by Giorgi [8]: "In introducing his rationalized, four-dimensional system, Giorgi [1901] wrote, *Il sistema CGS, con questo, perde ogni ragione di esistere; ma non credo che il suo abbandono sarà lamentato da alcuno*. ("With this, the CGS system loses every reason to exist; but I do not think that its abandonment will be lamented by anyone."). He may be correct, eventually." [7].

## 2. Unit conversion

### 2.1. Example of conversion miscalculation

Unit conversion should be a straightforward matter since there are tables available that provide instructions on how to properly convert the quantities from Gaussian & cgs-emu units to SI units. The most commonly

Table 1 Few entries from Goldfarb and Fickett table for the unit conversion between Gaussian & cgs-emu and SI. See ref. [9] for the full table and notes.

| Quantity | Symbol | Gaussian & cgs-emu[a] | Conversion factor, $C$[b] | SI |
|---|---|---|---|---|
| Magnetic flux density, magnetic induction | $B$ | gauss (G) | $10^{-4}$ | tesla(T), Wb/m$^2$ |
| (Volume) magnetization | $M$ | emu/cm$^3$ | $10^3$ | A/m |
| (Volume) magnetization | $4\pi M$ | G | $10^3/4\pi$ | A/m |
| Magnetic moment | $m$ | emu, erg/G | $10^{-3}$ | A·m$^2$ |

a. Gaussian and cgs-emu units are the same for magnetic properties. The defining relation is $B = H + 4\pi M$.
b. Multiply a number in Gaussian units by $C$ to convert it to SI (e.g., $1\,\text{G} \times 10^{-4}\,\text{T/G} = 10^{-4}\,\text{T}$).

used reference is probably the table compiled by Goldfarb and Fickett [9]. Unfortunately, this very useful table is sometimes misunderstood and improperly used, or its footnotes are neglected. Here, a few entries of this table (Table 1) are reported to explain its content, proper use, frequent misunderstandings and to show an instructive example of how the misunderstanding of a correct table could lead to macroscopic errors (only the entries necessary for this example are reported in Table 1).

To explain this example, a few items should be noted in this table. Firstly, there are two definitions for (volume) magnetization, which in Gaussian units is defined either as $M$, expressed in [emu/cm$^3$], or $4\pi M$, expressed in [G], as clearly explained in table note *a*. Both can be converted into SI [A/m], but with different conversion factors ($10^3$ and $10^3/4\pi$, respectively). As pointed out by Goldfarb [7], these two definitions of magnetization can generate confusion and possible mistakes. This is even more the case when considering old papers that use slightly different definitions of magnetization. The considered example is the paper of Gyorgy *et al.* [10] whose formula for calculating $J_c$ is widely used. It defines the magnetization hysteresis as $\Delta M$ in Gauss (i.e. the factor $4\pi$ is missing), which makes conversion even more confusing. $\Delta M$ in Gauss implies that this quantity has been determined from the relation $4\pi \Delta m [\text{emu}]/V[\text{cm}^3]$ and needs to be converted from [G] to [A/m] by using the second definition of magnetization in Table 1, which is $10^3/4\pi$ and not the conversion factor for the magnetic induction. Since the two $4\pi$ factors simplify, Gyorgy's formula is equivalent to a similar-looking formula where $\Delta M$ is expressed in [emu/cm$^3$], which requires the use of the first definition of magnetization in Table 1 and the conversion factor $10^3$. In cases like this, only a careful unit analysis and/or understanding of how the formula was derived allows identification of possible mismatches [11].

The second noteworthy detail in Table 1 is the footnote *b*, which explains that a given quantity in Gaussian units must be multiplied by the conversion factor $C$ to obtain its correct value in SI units. But, most importantly, it provides a conversion example, $1\,\text{G} \times 10^{-4}\,\text{T/G} = 10^{-4}\,\text{T}$, which is obviously an equality and explains that the conversion factors reported in the table are not dimensionless, but have units corresponding to the final units (SI) divided by the initial units (Gaussian & cgs-emu) (see for instance NIST best practices for unit conversion [12]).

*2.2. Dimensional conversion factors and table*

With the lesson learnt from the example above, a conversion table with explicit units in the conversion factors can be compiled (Table 2) on the basis of ref. [7]. Notice that since the two sides of the equation

Table 2 Table for unit conversion from Gaussian & cgs-emu to SI with explicit dimensional conversion factors (updated version of [1] and [2]). Multiply a number in Gaussian & cgs-emu units by $C$ to convert to SI units (e.g., 1 G ·$10^{-4}$ T/G = $10^{-4}$ T). Divide a number in SI units by $C$ to convert it to Gaussian & cgs-emu units (e.g., 1 T/($10^{-4}$ T/G) =$10^4$ G).

| Quantity | SI Symbol | Gaussian & cgs-emu[†] | Conversion factor, $C$ | SI |
|---|---|---|---|---|
| Magnetic flux density, magnetic induction | $B$ | G | $10^{-4}$ T/G | T (or Wb/m$^2$) |
| Magnetic flux | $\Phi$ | Mx, G·cm$^2$ | $10^{-8}$ Wb/Mx (or V·s/( G·cm$^2$)) | Wb, V·s |
| Permeability[*,§] | $\mu$ | dimensionless | $\{\mu_0\}$ H/m | H/m |
| Magnetic field strength, magnetizing force[§] | $H$ | Oe | $10^{-4}/\{\mu_0\}$ (A/m)/Oe | A/m |
| Magnetic moment | $m$ | emu, erg/G | $10^{-3}$ A·m$^2$/emu (or A·m$^2$/(erg/G)) | A·m$^2$ |
| Magnetic dipole moment[§] | $j$ | emu, erg/G | $10^{-3}\{\mu_0\}$ Wb·m/emu (or Wb·m/(erg/G)) | Wb·m |
| Magnetization[§] | $M$ | emu/cm$^3$, erg/(G·cm$^3$) | $10^3$ A/m/(emu/cm$^3$) | A/m |
|  |  | G | $10^{-4}/\{\mu_0\}$ (A/m)/G | A/m |
| Magnetic polarization, intensity of magnetization | $J, I$ | G | $10^{-4}$ T/G | T (or Wb/m$^2$) |
| (Mass) magnetization | $\sigma$ | emu/g | 1 A·m$^2$/kg/(emu/g) | A·m$^2$/kg |
| Susceptibility | $\chi$ | dimensionless | $4\pi$ | dimensionless |
| (Mass) susceptibility | $\chi_\rho, \chi_m$ | cm$^3$/g | $4\pi\cdot 10^{-3}$ m$^3$/kg/(cm$^3$/g) | m$^3$/kg |
| Energy density (volume) | $W$ | erg/cm$^3$ | $10^{-1}$ J/m$^3$/(erg/cm$^3$) | J/m$^3$ |
| Demagnetization factor | $N, D$ | dimensionless | $(4\pi)^{-1}$ | dimensionless |

[*] The SI relative permeability, $\mu_r = \mu/\mu_0 = 1+\chi$, corresponds to the emu permeability $\mu = 1+4\pi\chi$.
[†] Gaussian and cgs-emu units are the same for magnetic properties. The defining relation is $B = H + 4\pi M$.
[§] $\{\mu_0\}$ is the numerical value of the vacuum magnetic permeability $\mu_0$ after the 2019 revision of the SI system

must have identical units, the explicit units in the conversion factors $C$ makes obvious whether the quantities must be multiplied or divided by $C$. As already explained by Goldfarb [7], after the 2019 revision of the SI system $\mu_0$ does not have any longer the exact value of $4\pi\cdot 10^{-7}$, but is a measurable quantity subject to experimental error. In Table 2, the form $\{\mu_0\}$ refers to the numerical value of the vacuum magnetic permeability $\mu_0$, which is close to $4\pi\cdot 10^{-7}$ when measured in [H/m] units (or [Wb/A]) within a relative standard uncertainty of $1.6\cdot 10^{-10}$ (the present value is $\{\mu_0\} = (1.25663706127\pm 0.00000000020)\cdot 10^{-6}$).

Useful resources can be found also in [13], in particular for conversion of quantities or formulae between Gaussian and SI. It is important to notice that, despite the fact that [13] reports also conversion factors for natural units with $\hbar = c = 1$, these are valid for only one of the several widely-used natural-unit systems; sometimes also $k_B = 1$ is used, and the factors in [13] do not apply.

Vector magnetization is defined as $\boldsymbol{M} = d\boldsymbol{m}/dV$. The experimentally measured magnetic moment $m$ of a sample (which depends on the directions of mechanical movement and magnetic field) is the integral over the entire volume and the sample magnetization is defined as $M = m/V$ (both $\boldsymbol{M}$ and $M$ are measured in [A/m] in SI units, and in [emu/cm$^3$] or [erg/(G·cm$^3$)] in Gaussian & cgs-emu). However, in some circumstances (such as irregularly shaped or powdered samples) accurately determining the volume is difficult. In these situations, mass magnetization, $\sigma$, is often reported by normalizing the magnetic moment $m$ to the sample mass ($\sigma$ can also be expressed as $M/\rho$, with $\rho$ being the material density, emphasizing the possible impact of density variations). For this reason, the conversion factor for mass magnetization is also provided in Table 2. Although SI units are generally preferable, in the magnetization case Gaussian & cgs-emu units still offer the advantage of explicitly showing normalization to volume or mass through their units ([emu/cm$^3$] or [emu/g]). However, it is important to point out that, for superconductors, magnetization based on the volume should be used whenever possible (instead of mass magnetization) regardless of the unit system, because it enables the calculation of $\chi$, and so the evaluation of the shielded volume fraction in $m(T)$ characterizations. Another acceptable use of Gaussian & cgs-emu units is when very small magnetic induction $B$ is measured, making Gauss a more practical unit.

3. **Dimensional analysis to validate physical formulae and their conversion: tips and examples**

As mentioned above, dimensional analysis is an essential tool to verify and validate physical equations and, eventually, their interconversion. Here we provide a few basic rules and tips about dimensional analysis, and below some examples of how to use them.

In a valid physical equation:

- The units of added or subtracted terms must be the same.
- The two sides of the equation must have identical units.
- The argument of a function (unless raised to a power) must be dimensionless.

To ascertain dimensional homogeneity:

- In the paper of interest, verify whether the choice of the unit system is explicitly provided.
- If not, check the formulae for constants that are typical for specific unit system, like $c$ for the Gaussian system or $\mu_0$ for the SI system.
- Be cautious in using formulae provided in papers with mixed units.
- Some papers might provide values in SI units but relying on Gaussian & cgs-emu equations from older papers. If possible, verify the original paper.
- Check the formula for dimensional homogeneity following the above rules and, if possible, verify by other means for possible dimensionless parameters.

A few examples are useful to explain these concepts considering formulae and equations of particular interest for superconductivity.

*3.1. McMillan formula*

The first example can be found in the milestone paper by McMillan [3] for strongly-coupled superconductors. As in many theoretical papers of the time (and not only those) the constants are assumed to be equal 1 in all cases, resulting in equations such as:

$$\frac{T_\text{c}}{\omega_0} = \exp\left[\frac{-(1+\lambda)}{\lambda - \mu^* - (\langle\omega\rangle/\omega_0)\lambda\mu^*}\right] \quad (1)$$

and

$$T_c = \frac{\Theta}{1.45} \exp\left[-\frac{1.04(1+\lambda)}{\lambda - \mu^*(1+0.62\lambda)}\right] \quad (2)$$

where $\langle\omega\rangle$ and $\omega_0$ are defined as the average and maximum phonon frequency, $\Theta$ as the Debye characteristic phonon frequency, all of which are expressed in K ($\lambda$ and $\mu^*$ are the dimensionless electron-phonon coupling constant and Coulomb pseudopotential) [3]. These and similar formulae were used over the years either without specifying that the constants were assumed to be unity and/or expressing the values of the frequencies in different units (K, cm$^{-1}$, eV…) [14], [15], [16], [17], creating some difficulties in comparing values.

Equations (1) and (2) can be easily expressed as SI formulae by just paying attention to their dimensions. The right-hand side of eq. (1) is clearly dimensionless, implying that also the left-hand side must be turned into a dimensionless factor. Since in SI units $T_c$ is in K, and $\omega_0$ (which should be called an angular frequency; $\omega_0 = 2\pi\nu$, with $\nu$ being the frequency in s$^{-1}$) is in rad/s, the dimensional constants to be introduced are clearly $k_B$ and $\hbar$, leading to:

$$\frac{k_B T_c}{\hbar \omega_0} = \exp\left[\frac{-(1+\lambda)}{\lambda - \mu^* - (\langle\omega\rangle/\omega_0)\lambda\mu^*}\right] \quad \text{(SI formula)}$$

Eq. (2) could maintain the same form but replacing $\Theta$ with $\Theta_D$, the Debye temperature [18], [19]. The more frequently used Allen-Dynes modification [20] of the McMillan formula is reported with the proper constants for instance in ref. [21].

### 3.2. Ginzburg-Levanyuk number

Another example of useful dimension analysis can help to understand the correct formulation for the dimensionless Ginzburg-Levanyuk number, $Gi$, as mentioned above. An expression that can sometimes be found, without specifying the system, is:

$$Gi = \frac{1}{2}\left(\frac{8\pi^2 T_c \lambda^2}{\Phi_0^2 \xi}\right)^2 \quad (3)$$

We did, for instance, find this kind of expression with no explanation in a paper that was otherwise using SI units, but this is not an SI expression. Although the authors ultimately correctly calculated the $Gi$ numbers, mixing the system units is extremely misleading and can lead to errors. Eq. (3) can be derived, for instance, by the $Gi$ and $H_c$ expressions in ref. [22] (eqs. 2.47 and 2.7), which, however, uses Gaussian formulation with $k_B=1$. The unexpressed $k_B$ in eq. (3) can be found only by unit analysis because, without it, the quantity in parenthesis is in Ks$^2$/cm$^2$g=K/erg (in Gaussian & cgs-emu), which has the dimension of $1/k_B$, instead of being dimensionless as it must be. So, the Gaussian & cgs-emu formula with explicit constant for the $Gi$ number is:

$$Gi = \frac{1}{2}\left(\frac{8\pi^2 k_B T_c \lambda^2}{\Phi_0^2 \xi}\right)^2 \quad \text{(Gaussian \& cgs - emu formula)} \quad (4)$$

whereas the equivalent SI formula is (as already reported for instance in ref. [23]):

$$Gi = \frac{1}{2}\left(\frac{2\pi\mu_0 k_B T_c \lambda^2}{\Phi_0^2 \xi}\right)^2 \quad \text{(SI formula)} \quad (5)$$

The correctness of the conversion between eqs. (4) and (5) by using Table 2 and other simple conversions can be found in Appendix 1. In the same Appendix, we also demonstrate the use of the resources in [13].

*3.3. WHH equations*

A more complex example is in the classic Werthamer-Helfand-Hohenberg papers [6], [24] for the theory of the temperature-dependence of the upper critical field, $H_{c2}$. The following variables are defined in [6] as dimensionless:

$$\bar{h} = 2eH \left( \frac{v_F^2 \tau}{6\pi T_c} \right) \tag{6}$$

$$\alpha = \frac{3}{2mv_F^2 \tau} \tag{7}$$

$$\lambda_{so} = \frac{1}{3\pi T_c \tau_2} \tag{8}$$

where $\tau$ and $\tau_2$ are the transport and spin-orbital scattering times. However, these quantities do not appear dimensionless and, if reused without providing further explanations, these confusing notations can lead to misunderstanding and miscalculations. This is a perfect example of a case where the paper must be carefully read: in fact, the reason behind these apparently incorrect equations is that most of the equations in ref. [6] are expressed assuming that the fundamental constants $\hbar = c = k_B = 1$, as explained in section II (the paper uses otherwise Gaussian/cgs-emu formulation, as can be deduced by $H_{c2}$ values given in G or kG). However, proper constants are re-introduced in the last section where the variable $\alpha$ is expressed again as:

$$\alpha = \frac{3e^2 \hbar \gamma \rho_n}{2m\pi^2 k_B^2} \quad \text{(Gaussian \& cgs – emu formula)} \tag{9}$$

where $\gamma$ is the electronic specific-heat coefficient (Sommerfeld constant) and $\rho_n$ is the normal-state resistivity. Comparing eqs. (7) and (9) and taking into account the $\gamma$ and $\rho_n$ relations given for instance in [25]

$$\rho_n = \frac{1}{e^2 \tau} \left( \frac{m}{n} \right)$$

$$\gamma = \frac{\pi^2}{2} \left( \frac{k_B}{E_F} \right) n k_B = \frac{\pi^2 k_B^2}{v_F^2} \left( \frac{n}{m} \right)$$

allow us to verify also through this second form that the non-explicit constant in eq. (7) is indeed $\hbar$, leading to

$$\alpha = \frac{3\hbar}{2mv_F^2 \tau} \quad \text{(Gaussian \& cgs – emu and SI formula)} \tag{10},$$

as also reported in ref. [26].

Dimensional analysis makes it also clear that the non-explicit constants of eq. (8) are $\hbar/k_B$, so becoming

$$\lambda_{so} = \frac{\hbar}{3\pi k_B T_c \tau_2} \quad \text{(Gaussian \& cgs – emu and SI formula)} \tag{11}.$$

A little more attention needs to be paid to eq. (6), because it is not completely clear if ref. [6] uses Gaussian or cgs-emu. In the first case, the relation with explicit constants becomes

$$\bar{h} = \frac{2eH}{c}\left(\frac{v_F^2 \tau}{6\pi k_B T_c}\right) \quad \text{(Gaussian formula)},$$

whereas in the second case

$$\bar{h} = 2eH\left(\frac{v_F^2 \tau}{6\pi k_B T_c}\right) \quad \text{(cgs} - \text{emu formula)}.$$

However, because of the different definitions of the flux quantum ($\Phi_0 = hc/2e$ in Gaussian and $\Phi_0 = h/2e$ in cgs-emu), both can be written as:

$$\bar{h} = \frac{hH}{\Phi_0}\left(\frac{v_F^2 \tau}{6\pi k_B T_c}\right) \quad \text{(Gaussian \& cgs} - \text{emu formula)} \quad (12).$$

This is converted to SI as:

$$\bar{h} = \frac{h\mu_0 H}{\Phi_0}\left(\frac{v_F^2 \tau}{6\pi k_B T_c}\right) = 2e\mu_0 H\left(\frac{v_F^2 \tau}{6\pi k_B T_c}\right) \quad \text{(SI formula)} \quad (13).$$

It is trivial to verify that the Gaussian & cgs-emu and SI forms of eqs. (10) and (11) are identical. The conversion between eqs. (12) and (13) is shown in Appendix 2.

$\bar{h}$, $\alpha$ and $\lambda_{so}$ together with $t = T/T_c$ are then used in the key WHH equation that determines the temperature dependence of $H_{c2}$ [6]:

$$\ln\frac{1}{t} = \left(\frac{1}{2} + \frac{i\lambda_{so}}{4\bar{\gamma}}\right)\psi\left(\frac{1}{2} + \frac{\bar{h} + \frac{1}{2}\lambda_{so} + i\bar{\gamma}}{2t}\right) + \left(\frac{1}{2} - \frac{i\lambda_{so}}{4\bar{\gamma}}\right)\psi\left(\frac{1}{2} + \frac{\bar{h} + \frac{1}{2}\lambda_{so} - i\bar{\gamma}}{2t}\right) - \psi\left(\frac{1}{2}\right) \quad (14)$$

where $\psi$ is the digamma function and $\bar{\gamma} = \left[(\alpha\bar{h})^2 - \left(\frac{1}{2}\lambda_{so}\right)^2\right]^{1/2}$. It is important to notice, however, that according to the relations in the last section of [6], $\bar{h}$ can be expressed as $\bar{h} = \frac{4\,H_{c2}(T)}{\pi^2 T_c \left|\frac{dH_{c2}}{dT}\right|_{T_c}}$, and hence related to the slope of $H_{c2}$ at $T_c$, a parameter that is often more experimentally accessible. Since eq. (14) is expressed with dimensionless variables, it is independent of the unit system and the resulting $H_{c2}(T)$ curve will be in the same units of the analyzed data, but the variable definitions with proper form and units (like eqs. (10), (11) and (13) for SI) are needed to determine the quantities within (such as $v_F$). For $\alpha = \lambda_{so} = 0$, eq. (14) reduces to the simpler relation

$$\ln\frac{1}{t} = \psi\left(\frac{1}{2} + \frac{\bar{h}}{2t}\right) - \psi\left(\frac{1}{2}\right) \quad (15).$$

The 0 K limit of this equation is the well-known relation $H_{c2}(0) = (\pi^2/2)e^{\psi\left(\frac{1}{2}\right)}T_c\left|\frac{dH_{c2}}{dT}\right|_{T_c} \sim 0.693\, T_c\left|\frac{dH_{c2}}{dT}\right|_{T_c}$. In Appendix 3, we show how the reduced WHH formula in eq. (15) can be re-written in terms of the diffusivity $D$ or the $H_{c2}$ slope at $T_c$ and used to fit experimental data. We also describe the actual fitting function code provided in Supplementary material.

Notice that using explicit fundamental constants in the Introduction and Discussion sections, but assuming them equal to unity in the central part, is very common in theoretical papers also today. So particular attention must be paid to these articles, and it should always be stated clearly what unit system has been employed and if any constant has been assumed as equal to unity.

### 3.4. Bean model calculation of $J_{c,mag}$ in SI units and practical conversion

The Bean model calculation for a sample of rectangular cross-section is relatively simple. So, instead of converting the formula in [10], we calculate it directly. In SI units, the moment generated by a current circulating in a loop is given by $m = IA_{loop}$, where $I$ is the current in [A] and $A_{loop}$ is the area of the loop in [m$^2$] (notice that this equation is not valid in Gaussian units where $m = IA_{loop}/c$; in fact Gaussian and cgs-emu units are the same only for magnetic properties, as stated clearly in [9]). Assuming the simplest Bean scenario in which the field $B$ has fully penetrated the sample, the induced current in a sample of rectangular cross-section circulates in rectangular loops with the same absolute value of $J_{c,mag}$ at all points in the cross-section, as shown in Figure 1, and without change throughout the sample thickness $t$. Here we assume that the sample cross-section is $w \times l = 2a \times 2b$, with $a < b$. Considering the sketch in Figure 1 and its symmetry, the differential of the magnetic moment can be expressed as:

$$dm(x) = dIA_{loop} = (J_{c,mag}t dx)(4(b - a + x)x)$$

The total moment $m$ generated by the supercurrent can be calculated by integration as:

$$m = 4J_{c,mag}t \int_0^a (b - a + x)x \, dx = 4J_{c,mag}t \left[\frac{a^3}{3} + \frac{a^2}{2}(b - a)\right]$$

Since the sample volume is $V = 4abt$ and the magnetization is by definition $M = m/V$, we obtain:

$$M = \frac{J_{c,mag}a}{2}\left(1 - \frac{a}{3b}\right).$$

The amplitude of the hysteresis loop is $\Delta M = 2M$, so:

$$\Delta M = J_{c,mag}a\left(1 - \frac{a}{3b}\right),$$

as already reported for instance in ref. [27]. Substituting $a = w/2$ and $b = l/2$, we obtain the following SI formulae for $J_{c,mag}$:

$$J_{c,mag}\left[\frac{A}{m^2}\right] = \frac{2\Delta M\left[\frac{A}{m}\right]}{w[m]\left(1 - \frac{w[m]}{3l[m]}\right)} = \frac{2\Delta m[Am^2]}{V[m^3]w[m]\left(1 - \frac{w[m]}{3l[m]}\right)} \quad \text{(SI units)} \quad (16)$$

where $J_{c,mag}$ in [A/m$^2$], $\Delta M$ in [A/m], $w$ and $l$ (being the full cross-section dimensions with $w < l$) in [m], and $V$ in [m$^3$]. This, and similar equations, mean that only variables in the indicated units may be substituted to obtain $J_{c,mag}$ in the indicated units. In this case all variables, and the result, are in SI.

Since most instruments provide the moment in [emu], using eq. (16) requires first converting such [emu] moments to $m$ [Am$^2$] with the relation $\Delta m[Am^2] = 10^{-3} \frac{Am^2}{emu} \Delta m[emu]$ (Table 2), expressing $w$, $l$ and $t$ in [m] and $V$ in [m$^3$]. If desired, the final conversion from A/m$^2$ to A/cm$^2$ is trivial.

Despite SI units and formulae are nowadays preferable, a hybrid cgs-emu/SI Bean formula with the critical current density in A/cm², hysteresis moment in emu and spatial dimensions in cm is still acceptable, if it is properly used, understood in its meaning and applied to convert quantities properly. For this reason, such a formula is here derived, starting from the SI formula in eq. (16). Since eq. (16) requires SI variables, all unit conversions are made explicit (even when unnecessary):

$$\Delta m[\text{Am}^2] = 10^{-3} \frac{\text{Am}^2}{\text{emu}} \Delta m[\text{emu}] \quad (17)$$

$$w[\text{m}] = 10^{-2} \frac{\text{m}}{\text{cm}} w[\text{cm}] \quad (18)$$

$$V[\text{m}^3] = 10^{-6} \frac{\text{m}^3}{\text{cm}^3} V[\text{cm}^3] \quad (19)$$

$$J_{c,\text{mag}}\left[\frac{\text{A}}{\text{m}^2}\right] = 10^4 \frac{\text{cm}^2}{\text{m}^2} J_{c,\text{mag}}\left[\frac{\text{A}}{\text{cm}^2}\right] \quad (20)$$

Obtaining a cgs-emu/SI hybrid formula simply requires the substitution of eqs. (17)-(20) into eq. (16) (second form) without any further conversion. Writing explicitly successive simplification passages, we have:

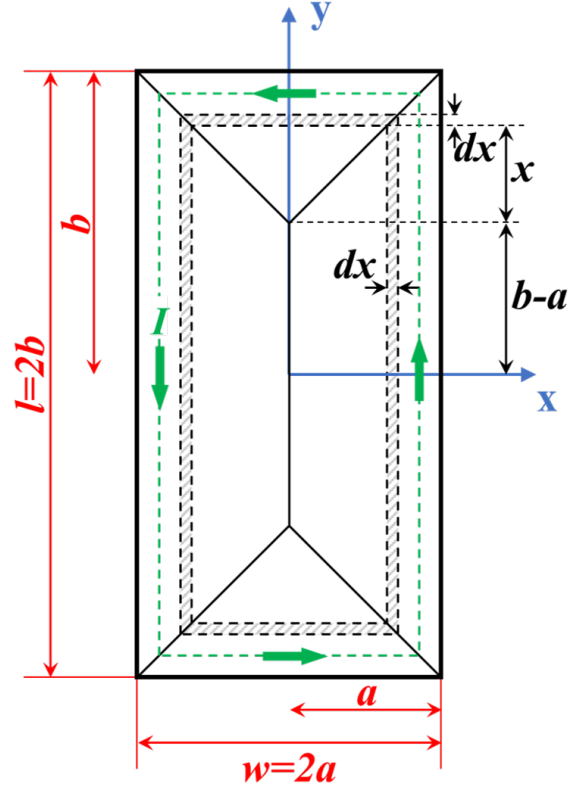

**Figure 1** Schematic diagram for the calculation of $J_{c,\text{mag}}$ in the simplest Bean model scenario. The field is perpendicular to the sample rectangular cross-section $w \times l = 2a \times 2b$ (with $a < b$) and the current circulates in rectangular loops with the same absolute value of $J_{c,\text{mag}}$ at all points in the cross-section.

$$10^4 \frac{\text{cm}^2}{\text{m}^2} J_{c,\text{mag}}\left[\frac{\text{A}}{\text{cm}^2}\right] = \frac{2 \cdot 10^{-3} \frac{\text{Am}^2}{\text{emu}} \Delta m[\text{emu}]}{10^{-6} \frac{\text{m}^3}{\text{cm}^3} V[\text{cm}^3] 10^{-2} \frac{\text{m}}{\text{cm}} w[\text{cm}] \left(1 - \frac{w[\text{cm}]}{3l[\text{cm}]}\right)}$$

$$J_{c,\text{mag}}\left[\frac{\text{A}}{\text{cm}^2}\right] = \frac{2 \cdot 10^{-3} \frac{\text{Am}^2}{\text{emu}} \Delta m[\text{emu}]}{10^4 \frac{\text{cm}^2}{\text{m}^2} 10^{-6} \frac{\text{m}^3}{\text{cm}^3} V[\text{cm}^3] 10^{-2} \frac{\text{m}}{\text{cm}} w[\text{cm}] \left(1 - \frac{w[\text{cm}]}{3l[\text{cm}]}\right)}$$

$$J_{c,\text{mag}}\left[\frac{\text{A}}{\text{cm}^2}\right] = \frac{20 \frac{\text{Acm}^2}{\text{emu}} \Delta m[\text{emu}]}{V[\text{cm}^3] w[\text{cm}] \left(1 - \frac{w[\text{cm}]}{3l[\text{cm}]}\right)} \quad \text{(hybrid cgs emu/SI units)} \quad (21)$$

which demonstrates that the pre-factor 20 in the commonly used practical hybrid Bean formula has indeed the dimension Acm²/emu (and is not just a pure number), as was also previously demonstrated by unit analysis in [28]. As stated above, the proper use of eq. (21) requires substituting the quantities in the indicated units to obtain $J_{c,\text{mag}}$ in the indicated units.

Similar calculation can also be done for the conversion of the first form of eq. (16) (see Appendix 4), leading to:

$$J_{c,\text{mag}}\left[\frac{A}{\text{cm}^2}\right] = \frac{20 \frac{A\text{cm}^2}{\text{emu}} \Delta M \left[\frac{\text{emu}}{\text{cm}^3}\right]}{w[\text{cm}]\left(1 - \frac{w[\text{cm}]}{3l[\text{cm}]}\right)} \quad \text{(hybrid cgs emu/SI units)} \quad (22)$$

Note that eqs. (21) and (22) are practical conversion formulae that use cgs-emu quantities as input data and an SI submultiple (A/cm²) as output, so they are neither cgs-emu nor SI formulae: in fact, Ampere is not the unit of current in cgs-emu, but abA (abampere) or Bi (biot) (in cgs-esu and Gaussian units, the current is expressed in statA or esu/s). These hybrid-unit formulae have clear potential dangers and should be used cautiously. It is important to point out that, unlike pure SI or cgs formulae, practical hybrid formulae like eqs. (21) and (22) should always have explicit units.

In all the equations given above, including the hybrid ones, dimensional homogeneity is verified (e.g. eq. (22) has A/cm² units of both sides of the equation).

Examples of the use of eqs. (16) and (21) are provided in Appendix 5.

### 4. Discussion

Superconductivity has a long history that encompasses many decades, during which time several unit systems have been employed. When considering theories and equations derived by others, it is always necessary to first verify the unit system used. Once identified, the original form can be used or proper conversion can be made but, in both cases, the unit system should be explicitly stated in any articles. We provide a few examples to demonstrate this approach. In particular, the best approach to calculate $J_{c,\text{mag}}$ from magnetic measurements is by using SI units and formulae throughout, making the cgs-emu to SI conversion of magnetic moment $m$ right at the start. This paper shows that if a converted equation is desirable, its validity must be verified against a known valid equation to avoid conversion errors. It is probably true that many mistakes have been made over the years and perhaps overlooked. It is not our intention here to identify specific examples. But, since conversion errors lead to values that are typically off by orders of magnitude, it is important to confirm the results by alternative methods before claiming world records. In the case of critical current density data, results reporting exceptional $J_c$ values should be independently verified and assessed, preferably with transport measurements, and ideally by independent research group(s). If only magnetic moment measurements were made, it is also highly advisable to provide the raw data, the equation used to convert $m$ to $J_c$, and all necessary information (such as sample sizes for $J_{c,\text{mag}}$ calculation) needed for a proper data analysis to allow the scientific community to gain a reliable assessment. Moreover, a direct comparison with a reference sample that does not show special performance, combined with valid physical explanations that justify sudden improvements in performance would be an effective approach to validate any extraordinary result. Without physical explanations, any reported records well exceeding those of similar samples should raise suspicion and would inevitably generate skepticism.

### 5. Conclusion

Although it might appear anachronistic to write about unit conversion and use of the Bean model in 2026, recent errors showed that it is still necessary and timely. Decades of articles provide invaluable resources for present-day research. However, many of these articles did not employ SI units and formulations, preferring Gaussian & cgs-emu or other systems or even neglecting constants altogether. To make valuable use of those resources, understanding their real meaning and converting them into today's recommended

SI notations is vital. Unit analysis is an effective tool to evaluate scientific formulae, and to derive correct unit conversion when necessary or desired. Here we provided tools and strategies for the conversion of magnetic quantities with explicit units in the conversion factors (Table 2), calculated the SI formula for the evaluation of $J_c$ by magnetization within the simplest Bean model scenario and verified the correctness of a frequently used hybrid-unit formula providing examples. Furthermore, we stated correct SI versions of the McMillan formula, the Ginzburg-Levanyuk number $Gi$, as well as relations of parameters in the WHH theory.

Finally, as advised by Giorgi [8] more than a century ago and pointed out by Goldfarb [7] in more recent years, the preferable and better-understood SI units and their immediate multiples and submultiples should be recommended since Gaussian & cgs-emu units are falling in disuse. We need to advocate now with the commercial magnetometer manufacturers to provide an SI units option in their instruments, and with publishers to recommend the use of SI units and formulae in their publications more strongly.

**Acknowledgements**

Part of this work was performed at the National High Magnetic Field Laboratory, which is supported by National Science Foundation Cooperative Agreements DMR-2128556, and by the State of Florida. The work of E.F.T. was carried out within the framework of the state assignment of the Ministry of Science and Higher Education of the Russian Federation for the IMP UB RAS. E.F.T. gratefully acknowledged the research funding from the Ministry of Science and Higher Education of the Russian Federation under Ural Federal University Program of Development within the Priority-2030 Program.

**ORCID iDs**

Chiara Tarantini https://orcid.org/0000-0002-3314-5906
Evgeny F. Talantsev https://orcid.org/0000-0001-8970-7982
Ilaria Pallecchi https://orcid.org/0000-0001-6819-6124
Jens Hänisch https://orcid.org/0000-0003-2757-236X
Jeffery L. Tallon https://orcid.org/0000-0002-4036-4429
David C. Larbalestier https://orcid.org/0000-0001-7098-7208

**Appendix 1**

The best way to convert the Gaussian & cgs-emu eq. (4) into SI is by writing down first all conversion factors with their units. Although the use of all units might appear cumbersome, it prevents mistakes. So, the conversion factors for all quantities in eq. (4) are:

$$\lambda[\text{cm}] = 10^2 \frac{\text{cm}}{\text{m}} \lambda[\text{m}]$$

$$\xi[\text{cm}] = 10^2 \frac{\text{cm}}{\text{m}} \xi[\text{m}]$$

$$k_B \left[\frac{\text{erg}}{\text{K}}\right] = 10^7 \frac{\text{erg}}{\text{J}} k_B \left[\frac{\text{J}}{\text{K}}\right]$$

$$\Phi_0[\text{Mx}] = 10^8 \frac{\text{Mx}}{\text{Wb}} \Phi_0[\text{Wb}]$$

Substituting all these quantities in eq. (4), we obtain:

$$Gi = \frac{1}{2}\left[\frac{8\pi^2 10^7 \frac{\text{erg}}{\text{J}} k_B \left[\frac{\text{J}}{\text{K}}\right] T_c[\text{K}] \left(10^2 \frac{\text{cm}}{\text{m}} \lambda[\text{m}]\right)^2}{\left(10^8 \frac{\text{Mx}}{\text{Wb}} \Phi_0[\text{Wb}]\right)^2 10^2 \frac{\text{cm}}{\text{m}} \xi[\text{m}]}\right]^2 = \frac{1}{2}\left[\frac{8\pi^2 10^{11} \frac{\text{gcm}^2 s^{-2}}{\text{J}} \frac{\text{cm}}{\text{m}} k_B \left[\frac{\text{J}}{\text{K}}\right] T_c[\text{K}](\lambda[\text{m}])^2}{10^{18} \frac{\text{gcm}^3 s^{-2}}{\text{Wb}^2} (\Phi_0[\text{Wb}])^2 \xi[\text{m}]}\right]^2$$

$$= \frac{1}{2}\left[\frac{8\pi^2}{10^7} \frac{\text{Wb}^2 \text{A}}{\text{JmA}} \frac{k_B \left[\frac{\text{J}}{\text{K}}\right] T_c[\text{K}](\lambda[\text{m}])^2}{(\Phi_0[\text{Wb}])^2 \xi[\text{m}]}\right]^2 = \frac{1}{2}\left[2\pi\mu_0 \left[\frac{\text{Wb}}{\text{A}}\right] \frac{\text{WbA}}{\text{Jm}} \frac{k_B \left[\frac{\text{J}}{\text{K}}\right] T_c[\text{K}](\lambda[\text{m}])^2}{(\Phi_0[\text{Wb}])^2 \xi[\text{m}]}\right]^2$$

where we used the equalities erg=gcm²s⁻², Mx=cm³ᐟ²g¹ᐟ²s⁻¹ and in the last step $\mu_0 = 4\pi/10^7$ Wb/A. Since all units simplify, the last step leads to a dimensionless quantity whose SI formula is:

$$Gi = \frac{1}{2}\left(\frac{2\pi\mu_0 k_B T_c \lambda^2}{\Phi_0^2 \xi}\right)^2 \qquad \text{(SI formula)}$$

Formulae can be converted also using the approach in [13] (section "Formula Conversion"), which is more compact but does not use explicit units. There are two factors defined, $\alpha = 10^2$ cm m⁻¹ and $\beta = 10^7$ erg J⁻¹, and the SI constants $\mu_0$, $\epsilon_0$ and $c = (\epsilon_0 \mu_0)^{-1/2}$ are used. A Gaussian quantity $\bar{Q}$ is converted into an SI quantity $Q$ through the relation $\bar{Q} = \bar{k} Q$ where $\bar{k}$ is the conversion factor. To convert the Gaussian relation:

$$\overline{Gi} = \frac{1}{2}\left(\frac{8\pi^2 \overline{k_B} T_c \bar{\lambda}^2}{\overline{\Phi_0}^2 \bar{\xi}}\right)^2$$

we need to use the following relations obtained from the table in [13] or derived from it:

$$\bar{\lambda} = \alpha \lambda$$

$$\bar{\xi} = \alpha \xi$$

$$\overline{k_B} = \beta k_B$$

$$\overline{\Phi_0} = \left(\frac{4\pi\beta}{\alpha^3 \mu_0}\right)^{1/2} \alpha^2 \Phi_0$$

Therefore, the SI relation for $Gi$ becomes:

$$Gi = \frac{1}{2}\left(\frac{8\pi^2 \beta k_B T_c \alpha^2 \lambda^2}{\left(\frac{4\pi\beta}{\alpha^3 \mu_0}\right) \alpha^4 \Phi_0^2 \alpha\xi}\right)^2 = \frac{1}{2}\left(\frac{2\pi\mu_0 k_B T_c \lambda^2}{\Phi_0^2 \xi}\right)^2$$

**Appendix 2**

Substituting

$$H[\text{Oe}] = \{\mu_0\} 10^4 \frac{\text{Oe}}{\text{A}/\text{m}} H\left[\frac{\text{A}}{\text{m}}\right] = \frac{4\pi}{10^3} \frac{\text{Oe}}{\text{A}/\text{m}} H\left[\frac{\text{A}}{\text{m}}\right]$$

$$h[\text{erg}\cdot\text{s}] = 10^7 \frac{\text{erg}}{\text{J}} h[\text{J}\cdot\text{s}]$$

$$v_F\left[\frac{\text{cm}}{\text{s}}\right] = 10^2 \frac{\text{cm}}{\text{m}} v_F\left[\frac{\text{m}}{\text{s}}\right]$$

and the conversions of $\Phi_0$ and $k_B$ (given in Appendix 1) in eq. (12), we obtain

$$\bar{h} = \frac{10^7 \frac{\text{erg}}{\text{J}} h[\text{J}\cdot\text{s}] \frac{4\pi}{10^3} \frac{\text{Oe}}{\text{A/m}} H\left[\frac{\text{A}}{\text{m}}\right]}{10^8 \frac{\text{Mx}}{\text{Wb}} \Phi_0[\text{Wb}]} \left(\frac{\left(10^2 \frac{\text{cm}}{\text{m}} v_F\left[\frac{\text{m}}{\text{s}}\right]\right)^2 \tau[\text{s}]}{6\pi 10^7 \frac{\text{erg}}{\text{J}} k_B\left[\frac{\text{J}}{\text{K}}\right] T_c[\text{K}]}\right)$$

$$= \frac{h[\text{J}\cdot\text{s}] \frac{4\pi}{10^3} \frac{\text{cm}^{-1/2}\text{g}^{1/2}\text{s}^{-1}}{\text{A/m}} H\left[\frac{\text{A}}{\text{m}}\right]}{10^8 \frac{\text{cm}^{3/2}\text{g}^{1/2}\text{s}^{-1}}{\text{Wb}} \Phi_0[\text{Wb}]} \left(10^4 \frac{\text{cm}^2}{\text{m}^2} \frac{\left(v_F\left[\frac{\text{m}}{\text{s}}\right]\right)^2 \tau[\text{s}]}{6\pi k_B\left[\frac{\text{J}}{\text{K}}\right] T_c[\text{K}]}\right)$$

$$= \frac{4\pi}{10^7} \frac{\text{Wb}}{\text{A/m}} \frac{1}{\text{m}^2} \frac{h[\text{J}\cdot\text{s}] H\left[\frac{\text{A}}{\text{m}}\right]}{\Phi_0[\text{Wb}]} \left(\frac{\left(v_F\left[\frac{\text{m}}{\text{s}}\right]\right)^2 \tau[\text{s}]}{6\pi k_B\left[\frac{\text{J}}{\text{K}}\right] T_c[\text{K}]}\right) =$$

$$= \mu_0\left[\frac{\text{Wb}}{\text{A}}\right] \frac{1}{\text{m}} \frac{h[\text{J}\cdot\text{s}] H\left[\frac{\text{A}}{\text{m}}\right]}{\Phi_0[\text{Wb}]} \left(\frac{\left(v_F\left[\frac{\text{m}}{\text{s}}\right]\right)^2 \tau[\text{s}]}{6\pi k_B\left[\frac{\text{J}}{\text{K}}\right] T_c[\text{K}]}\right)$$

where we used Oe = $\text{cm}^{-1/2}\text{g}^{1/2}\text{s}^{-1}$ (note that G = $\text{cm}^{-1/2}\text{g}^{1/2}\text{s}^{-1}$ too, so Oe and G have the same base units).

Since all units simplify in the above equation, this leads to a dimensionless variable whose SI formula is:

$$\bar{h} = \frac{h\mu_0 H}{\Phi_0}\left(\frac{v_F^2 \tau}{6\pi k_B T_c}\right) = 2e\mu_0 H\left(\frac{v_F^2 \tau}{6\pi k_B T_c}\right) \quad \text{(SI formula)}$$

where $\Phi_0 = h/2e$ in SI units.

**Appendix 3**

Expressing the variables in eq. (15) (or eq. (14)) in terms of diffusivity (or diffusion coefficient) $D$ can be advantageous because the Fermi velocity $v_F$ and the transport scattering time $\tau$ cannot be derived independently from each other solely on the basis of upper critical field data (see ref.[26]). In both SI and Gaussian units, the diffusion coefficient is defined as $D = \frac{v_F^2 \tau}{3}$ (by using the proper units for the two cases) with the following conversion

$$D\left[\frac{\text{cm}^2}{\text{s}}\right] = 10^4 \left(\frac{\text{cm}}{\text{m}}\right)^2 \cdot D\left[\frac{\text{m}^2}{\text{s}}\right]$$

According to the definition of $D$, eqs. (12) and (13) can be rewritten as:

$$\bar{h} = \frac{\hbar\mu_0 H D}{\Phi_0 k_B T_c} = \frac{4H}{\pi^2 \left|\frac{dH_{c2}}{dT}\right|_{T_c} T_c} \quad \text{(SI formula)} \tag{A1}$$

$$\bar{h} = \frac{\hbar H D}{\Phi_0 k_B T_c} = \frac{4H}{\pi^2 \left|\frac{dH_{c2}}{dT}\right|_{T_c} T_c} \quad \text{(Gaussian formula)} \tag{A2}$$

The second forms of eqs. (A1) and (A2) are due to relations between the absolute value of the $H_{c2}$ slope near $T_c$ and the diffusivity $D$, that in SI units is $\mu_0 \left|\frac{dH_{c2}}{dT}\right|_{T_c} = \frac{4}{\pi^2}\frac{\Phi_0 k_B}{\hbar D}$ and in Gaussian units $\left|\frac{dH_{c2}}{dT}\right|_{T_c} = \frac{4}{\pi^2}\frac{\Phi_0 k_B}{\hbar D}$ [29]. Therefore, we can write a single fitting equation, which by itself is independent of the units, in terms of the $H_{c2}$ slope in place of eq. (15):

$$\ln\frac{T}{T_c} + \psi\left(\frac{1}{2} + \frac{2H}{\pi^2 \left|\frac{dH_{c2}}{dT}\right|_{T_c} T}\right) - \psi\left(\frac{1}{2}\right) = 0 \tag{A3}$$

Here, we give details of the SI and Gaussian codes we provide in the Supplementary material, and below we will illustrate their use.

In the Supplementary material, we provide two Originlab fitting functions (requiring the Pro version because an implicit function must be resolved) based on the same eq. (A3) but differing in the definition of the fitting and derived parameters, which are evaluated in either SI or Gaussian units. For the SI version of the fitting function, the fitting parameters are "Tc_K" and "Slope_T_K" in [K] and [T/K] for $T_c$ and $\mu_0 \left|\frac{dH_{c2}}{dT}\right|_{T_c}$, while the derived parameters are "mu0Hc20_Tesla" and "Diffusivity_m2_s" in [T] and [m$^2$/s] for $\mu_0 H_{c2}(0)$ and $D$. For the Gaussian version, the fitting parameters are "Tc_K" and "Slope_Oe_K" in [K] and [Oe/K] for $T_c$ and $\left|\frac{dH_{c2}}{dT}\right|_{T_c}$, while the derived parameters are "Hc20_Oe" and "Diffusivity_cm2_s" in [Oe] and [cm$^2$/s] for $H_{c2}(0)$ and $D$. These functions are useful because they do not simply provide a fit to the experimental data, but also automatically calculate the most relevant physical parameters (note that

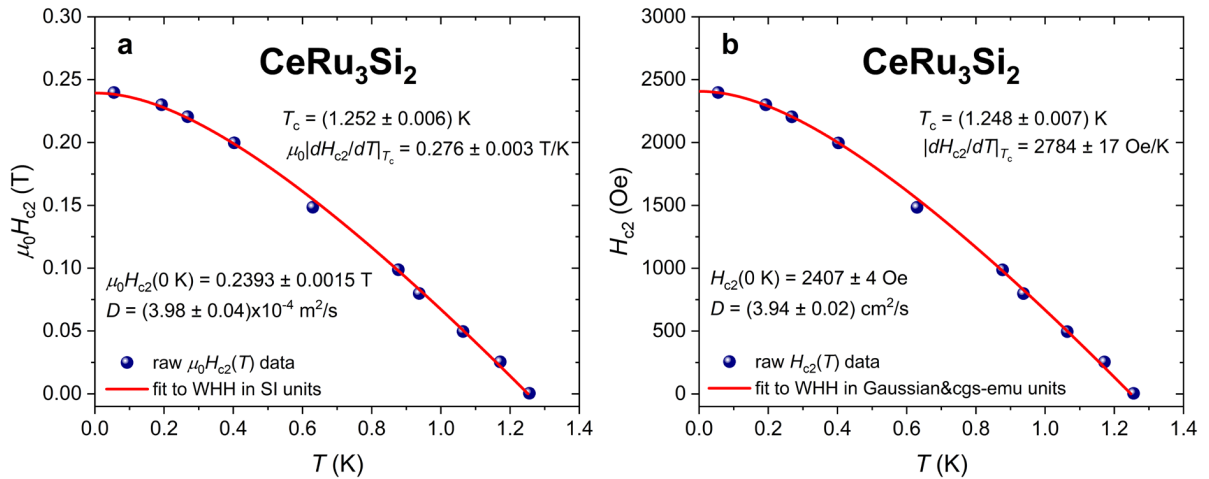

**Figure A1** Upper critical field data for CeRu$_3$Se$_2$ reported by Rauchschwalbe *et al.* [21] and fits to the reduced WHH model (eq. (A6)). (**a**) Data in the SI units and fit using the SI version of the provided fitting function; (**b**) Data in the Gaussian units and fit using the Gaussian version of the provided fitting function. In both cases the fit quality is better than R$^2$ = 0.99996.

Originlab automatically saves the derived parameters in the analysis report worksheet; those parameters can also be added in the report table on the graph by using "Quantities in Table").

For the demonstration we analyze the upper critical field data measured by Rauchschwalbe *et al.* [30] on a conventional weak-coupling (BCS) superconductor $CeRu_3Si_2$ (Fig. A1). The fits as well as the fitting and derived parameters are reported in the figure, showing compatible results taking into account the different units (the minor difference in the parameters obtained from the two fits is most likely caused by the approximation in calculating the argument of the digamma function $\psi$).

**Appendix 4**

Since

$$\Delta M \left[\frac{A}{m}\right] = 10^3 \frac{A \cdot cm^3}{m \cdot emu} \Delta M \left[\frac{emu}{cm^3}\right], \qquad (A4)$$

substituting eqs. (18)-(20) and (A4) into eq. (16) (first form) leads to:

$$10^4 \frac{cm^2}{m^2} J_{c,mag}\left[\frac{A}{cm^2}\right] = \frac{2 \cdot 10^3 \frac{A \cdot cm^3}{m \cdot emu} \Delta M \left[\frac{emu}{cm^3}\right]}{10^{-2} \frac{m}{cm} w[cm] \left(1 - \frac{w[cm]}{3l[cm]}\right)}$$

$$J_{c,mag}\left[\frac{A}{cm^2}\right] = \frac{2 \cdot 10^3 \frac{A \cdot cm^3}{m \cdot emu} \Delta M \left[\frac{emu}{cm^3}\right]}{10^4 \frac{cm^2}{m^2} 10^{-2} \frac{m}{cm} w[cm] \left(1 - \frac{w[cm]}{3l[cm]}\right)}$$

$$J_{c,mag}\left[\frac{A}{cm^2}\right] = \frac{20 \frac{Acm^2}{emu} \Delta M \left[\frac{emu}{cm^3}\right]}{w[cm] \left(1 - \frac{w[cm]}{3l[cm]}\right)} \qquad (A5)$$

**Appendix 5**

Let's assume that a magnetic hysteresis of 1.5 emu is measured in a sample with a cross-section of 4 mm by 5 mm and a thickness of 700 nm, and we want to calculate $J_{c,mag}$.

1) If we want to use the SI formula, eq. (16), the first step is to convert all variables into SI, so:

$$\Delta m[Am^2] = 10^{-3} \frac{Am^2}{emu} \Delta m[emu] = 10^{-3} \frac{Am^2}{emu} \cdot 1.5 \text{ emu} = 1.5 \cdot 10^{-3} \text{ Am}^2$$

$$w[m] = 4 \cdot 10^{-3} \text{ m}$$

$$l[m] = 5 \cdot 10^{-3} \text{ m}$$

$$t[m] = 7 \cdot 10^{-7} \text{ m}$$

$$V[m^3] = w[m] \times l[m] \times t[m] = 1.4 \cdot 10^{-11} \text{ m}^3$$

Substituting these values directly into eq. (16) in one single step without any further conversion, we obtain:

$$J_{c,\text{mag}}\left[\frac{A}{m^2}\right] = \frac{2\Delta m[Am^2]}{V[m^3]w[m]\left(1 - \frac{w[m]}{3l[m]}\right)} = \frac{2 \cdot 1.5 \cdot 10^{-3}\ \text{A}\cancel{m^2}}{1.4 \cdot 10^{-11}\ \text{m}^{\cancel{3}} \cdot 4 \cdot 10^{-3}\text{m}\left(1 - \frac{4 \cdot 10^{-3}\ \cancel{m}}{3 \cdot 5 \cdot 10^{-3}\ \cancel{m}}\right)}$$

$$= \frac{3}{5.6\left(\frac{11}{15}\right)} \cdot 10^{11}\ \frac{A}{m^2} \simeq 7.3\ 10^{10}\ \frac{A}{m^2} \tag{A6}$$

2) If we want to use the practical hybrid formula, eq. (21), we need to convert all variables in the units this formula requires, so:

$$\Delta m[\text{emu}] = 1.5\ \text{emu}$$

$$w[\text{cm}] = 4 \cdot 10^{-1}\ \text{cm}$$

$$l[\text{cm}] = 5 \cdot 10^{-1}\ \text{cm}$$

$$t[\text{cm}] = 7 \cdot 10^{-5}\ \text{cm}$$

$$V[\text{cm}^3] = w[\text{cm}] \times l[\text{cm}] \times t[\text{cm}] = 1.4 \cdot 10^{-5}\ \text{cm}^3$$

Substituting these values directly into eq. (21) in one single step without any further conversion, we obtain:

$$J_{c,\text{mag}}\left[\frac{A}{\text{cm}^2}\right] = \frac{20\ \frac{A\text{cm}^2}{\text{emu}}\Delta m[\text{emu}]}{V[\text{cm}^3]w[\text{cm}]\left(1 - \frac{w[\text{cm}]}{3l[\text{cm}]}\right)} = \frac{20\ \frac{A\cancel{\text{cm}^2}}{\cancel{\text{emu}}} \cdot 1.5\ \cancel{\text{emu}}}{1.4 \cdot 10^{-5}\ \text{cm}^{\cancel{3}} \cdot 4 \cdot 10^{-1}\text{cm}\left(1 - \frac{4 \cdot 10^{-1}\ \cancel{\text{cm}}}{3 \cdot 5 \cdot 10^{-1}\ \cancel{\text{cm}}}\right)}$$

$$= \frac{3}{5.6\left(\frac{11}{15}\right)} \cdot 10^{7}\ \frac{A}{\text{cm}^2} \simeq 7.3\ 10^{6}\ \frac{A}{\text{cm}^2}$$

This value, accordingly to eq. (20), matches the $J_{c,\text{mag}}$ calculated with the SI formula in eq.(A6) and in both cases dimensional homogeneity is preserved.